\documentclass[aps,reprint,twocolumn,amsmath]{revtex4-1}
\usepackage{CJK}
\usepackage{graphicx,color}
\usepackage{lipsum}
\usepackage{bm}
\begin{document}
		\title{Disentangling boson peaks and Van Hove singularities in a model glass}
		
		\author{Yinqiao Wang$^1$, Liang Hong$^{1,2}$, Yujie Wang$^1$, Walter Schirmacher$^3$ and Jie Zhang$^{1,2,4}$}
		\email{Electronic address: jiezhang2012@sjtu.edu.cn}
		\affiliation{$^1$School of Physics and Astronomy, Shanghai Jiao Tong University, 800 Dong Chuan Road, Shanghai 200240, China}
		\affiliation{$^2$Institute of Natural Sciences, Shanghai Jiao Tong University, Shanghai 200240, China}
		\affiliation{$^3$Institut f$\ddot{u}$r Physik, Universit$\ddot{a}$t Mainz, Staudinger Weg 7, D-55099 Mainz, Germany}
		\affiliation{$^4$Collaborative Innovation Center of Advanced Microstructures, Nanjing 210093, China}

\begin{abstract}

Using the example of a two-dimensional macroscopic model glass in which the interparticle forces can be precisely measured,
we obtain strong hints for resolving a controversy concerning the origin of the anomalous enhancement of the vibrational spectrum in glasses (boson peak).
Whereas many authors attribute this anomaly to the structural disorder, some other authors claim that the short-range order, leading to washed-out Van Hove singularities, would cause the boson-peak anomaly.
As in our model system, the disorder-induced and short-range-order-induced features can be completely separated, we are able to discuss the controversy about the boson peak in real glasses in a new light.
Our findings suggest that the interpretation of the boson peak in terms of short-range order only, might result from a coincidence
of the two phenomena in the materials studied. In general, as we show, the two phenomena both exist, but are two completely
separate entities. 

\end{abstract}

\maketitle

\section{introduction}
Glass shows a deviation in its vibrational density of states (DOS) from Debye's $\omega^{d-1}$ law, where $d$ is the dimensionality, which occurs in the THz regime, about one-tenth of the Debye frequency $\omega_D$.
This deviation leads to a peak  in the reduced DOS, $g (\omega)/\omega^{d-1}$ [boson peak (BP)] \cite{Buchenau86,Sokolov86,chumakov04}.
The origin of the BP is still under intense debate. The main controversy is, whether it is the result of the structural {\it disorder} \cite{Karpov83,Elliott92,Gurevich93,Schirmacher98,Schirmacher06,Schirmacher07,Marruzzo13}, or the glassy counterpart of the first (transverse) Van Hove singularity (VHS) in crystals \cite{Chumakov11,Chumakov14,Chumakov15,Chumakov16}, i.e. the result
of the {\it short-range order} of the glass.

In their recent publications about a glassy mineral and glassy SiO$_2$, Chumakov et al. \cite{Chumakov11,Chumakov14}
compared the DOS and the specific heat of the glassy materials with the spectra of the corresponding crystalline materials.
They found that the BP frequency of glass -- if rescaled to the corresponding crystalline density-coincides with the position of the first (transverse) VHS of the corresponding crystal. This was also substantiated for other materials \cite{Chumakov15}.
From this coincidence, they concluded that the BP would be the same physical phenomenon as the VHS in the crystal, namely coming from the piling up of resonances as a result of the bending down of the phonon dispersion near the edge of the pseudo Brillouin-zone (BZ) $k_p=k_0/2\approx\pi/a$ ($k_0$ is the wavenumber of the first sharp diffraction peak and $a$ is a mean intermolecular spacing) \cite{Chumakov16}.
It is possible to reformulate this point of view in terms of length scales: if the wavelength becomes short enough,
the wave is sensitive to the atomic order (or the short-range order in glass) so that the dispersion
bends down and leads to the VHS.

On the other hand, there is ample evidence from experimental \cite{MonacoGiordano09,Baldi10} and numerical \cite{MonacoMossa09,Marruzzo13} work that the BP in glass is associated with a disorder-induced rapid increase of the
Brillouin line width of the transverse excitation and a characteristic dip in the transverse sound velocity. 
These anomalies have been shown to result from the disorder in the elastic constants (elastic heterogeneity \cite{Duval90,Leonforte06,Tsamados09,Marruzzo13,Schirmacher15}).
It has been demonstrated that all these BP-related anomalies occur, because the wavelengths of the acoustic excitations get small enough to be sensitive to the breakdown of the translational, rotational and inversion symmetries \cite{Elliott92,Leonforte06,Tsamados09,MonacoGiordano09,Milkus16}.
As fluctuations of the shear modulus around a rather small value imply the existence of ``soft spots'', where the limit of structural stability is reached, this view of the BP origin is also consistent with the soft-potential model \cite{Karpov83,Buchenau86,Buchenau92,Parshin93,Gurevich93} and the view of the vicinity of a saddle transition \cite{Grigera03}.

So the controversy between the two views is whether the BP occurs as a result of short-range {\it order} or as a result of structural {\it disorder.} In 3D structural glasses, the length scales, where these local features become distinct, are not very different, so one cannot clearly distinguish between the two aspects.

In our experiment, we found that the two length scales - and correspondingly the two characteristic frequencies - are clearly separate, showing that the first VHS and the BP are two separate entities.

\section{Experimental methods}
\begin{figure}
	\centerline{\includegraphics[trim=0cm 0cm 0cm 0cm, width=1\linewidth]{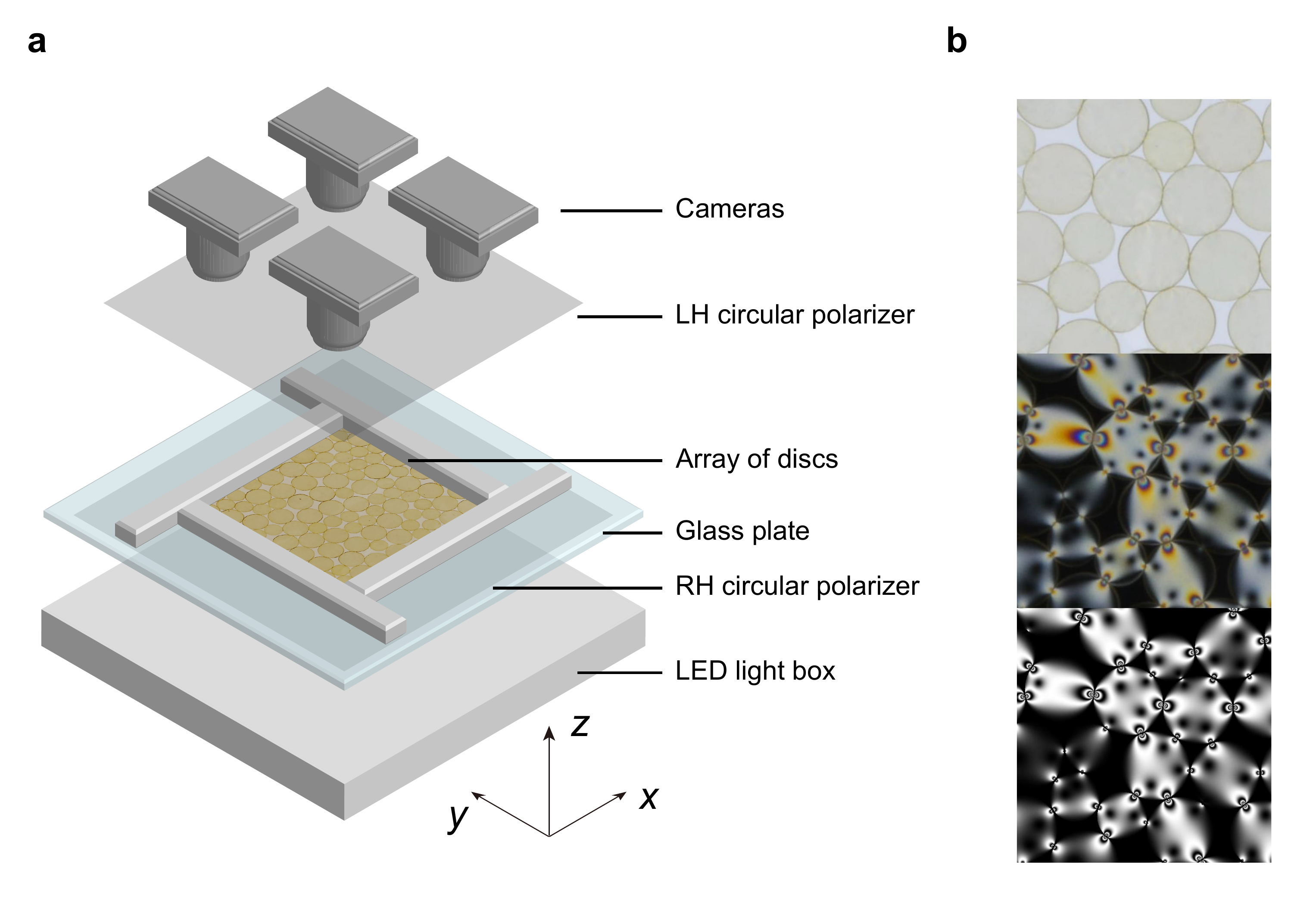}}
	\caption{a, Schematic of the experimental setup. b, Top: normal image without polarizers. Middle: stress image with polarizers. Bottom: reconstructed stress image based on calculated contact forces.}
	\label{Fig1}
\end{figure}

In this experiment (see Fig. \ref{Fig1}), we used a biaxial apparatus (or simply ``biax'') \cite{Majmudar05,Zhang17} to prepare an isotropically compressed jammed packing of photo-elastic disks. Viewed from the above, the biax consisted of a square domain with four mobile walls, whose positions could be precisely controlled with an accuracy of $0.1 \rm mm$ using Panasonic servo motors to move symmetrically when applying isotropic compression.

We filled the square domain with 2720 large disks ($D_L=1.4 \rm cm$) and 2720 small disks ($D_S=1.0 \rm cm$). These disks were randomly deposited to maximize the mixing of two types of disks.
The biax was mounted on a glass plate, on top of which the two-dimensional horizontal disk layer was placed. The surface between glass plate and disks was powder lubricated to minimize friction.
Viewed from the side, below the glass plate, a circular polarizer sheet was attached. Below this sheet, an LED light source provided uniform illumination of the disk layer. One and a half meters above the biax, an array of $2\times2$ high resolution cameras were mounted to take images of the whole disk packing. Right below the cameras, a second (matched) circular polarizer sheet was mounted horizontally, which could be freely inserted or removed so that two types of images of disk packing were taken to record disk configurations and stress information.

Since the total disk number was fixed, the packing fraction (the total disk area over the area of the square domain) was essentially determined by the size of the square domain, as controlled by the biax. In preparing the jammed packing, we applied gentle vibrations to the disk layer to break transient force chains due to the friction between disks to achieve a homogeneous and stress-free state before the packing fraction exceeded $\sim84\%$, which is the typical value of the isotropic jamming point of bi-disperse frictionless disks. 

We estimate that the contribution of elastic energy due to tangential contact forces only amounts to a few percent of the total elastic energy of the system. The data presented in the main text came from one packing configuration, while several other configurations were prepared using the same protocol. The differences of the data in the DOS and related properties between different configurations are slight, comparable to the symbol sizes in the figures.

The forces between these disks can be accurately determined (see Refs. \cite{Majmudar05,Zhang17}).
We applied imaging processing to extract the spring constants $k_n$ and $k_t$ at individual contacts using the calibrated contact-force laws and the values of contact forces. From these quantities we constructed the harmonic dynamical matrix $H_{ij}$ (Hessian) as follows:
\begin{widetext}
	\begin{eqnarray*}
		H_{ij}=\frac{1}{\sqrt{m_i m_j}}
		\begin{bmatrix}
			&k_{ij,n}\cos^2\theta_{ij}+(k_{ij,t}-f_{ij,n}/r_{ij})\sin^2 \theta_{ij}
			&(k_{ij,n}-k_{ij,t}+f_{ij,n}/r_{ij})\cos \theta_{ij}\sin \theta_{ij} \\
			&(k_{ij,n}-k_{ij,t}+f_{ij,n}/r_{ij})\cos \theta_{ij}\sin \theta_{ij}   &k_{ij,n}\sin^2\theta_{ij}+(k_{ij,t}-f_{ij,n}/r_{ij})\cos^2 \theta_{ij}
		\end{bmatrix}
		,\ i\neq j\,\, ,
	\end{eqnarray*}
\end{widetext}
$m_i$ is the mass of a disk
$i$, $k_{ij,n}$ are the normal,
$k_{ij,t}$ the tangential spring constants between disk $i$ and disk $j$.
$\theta_{ij}$ is the orientation angle of the bond between disk $i$ and disk $j$.
$f_{ij,n}$ is the normal force between disk $i$ and disk $j$.
$r_{ij}$ is the length of the bond between disk $i$ and disk $j$.
The matrix elements of the Hessian obey $H_{ii}=-\sum_{j\neq i}H_{ij}$.

\section{Results}
\subsection{The boson peak and Van Hove singularities}
By diagonalizing $H_{ij}$, we obtained the eigenvalues $\omega_\lambda^2$, and the eigenvectors $\mathbf{e}_\lambda
=\{\mathbf{e}_\lambda(1),\mathbf{e}_\lambda(2),\dots\mathbf{e}_\lambda(N)\}$.
From these $\omega_\lambda$ we obtained the DOS
\begin{equation}\label{dos}
g(\omega)=
\frac{1}{2N}\sum_{\lambda}\delta(\omega-\omega_{\lambda})
\end{equation}
and the single-site DOS \cite{Shintani08}
\begin{equation}\label{dosi}
g_i(\omega)=\frac{1}{2N}\sum_{\lambda}
|\mathbf{e}_\lambda(i)|^2
\delta(\omega-\omega_{\lambda})\, .
\end{equation}

\begin{figure}
	\centerline{\includegraphics[trim=0cm 0cm 0cm 0cm, width=.8\linewidth]{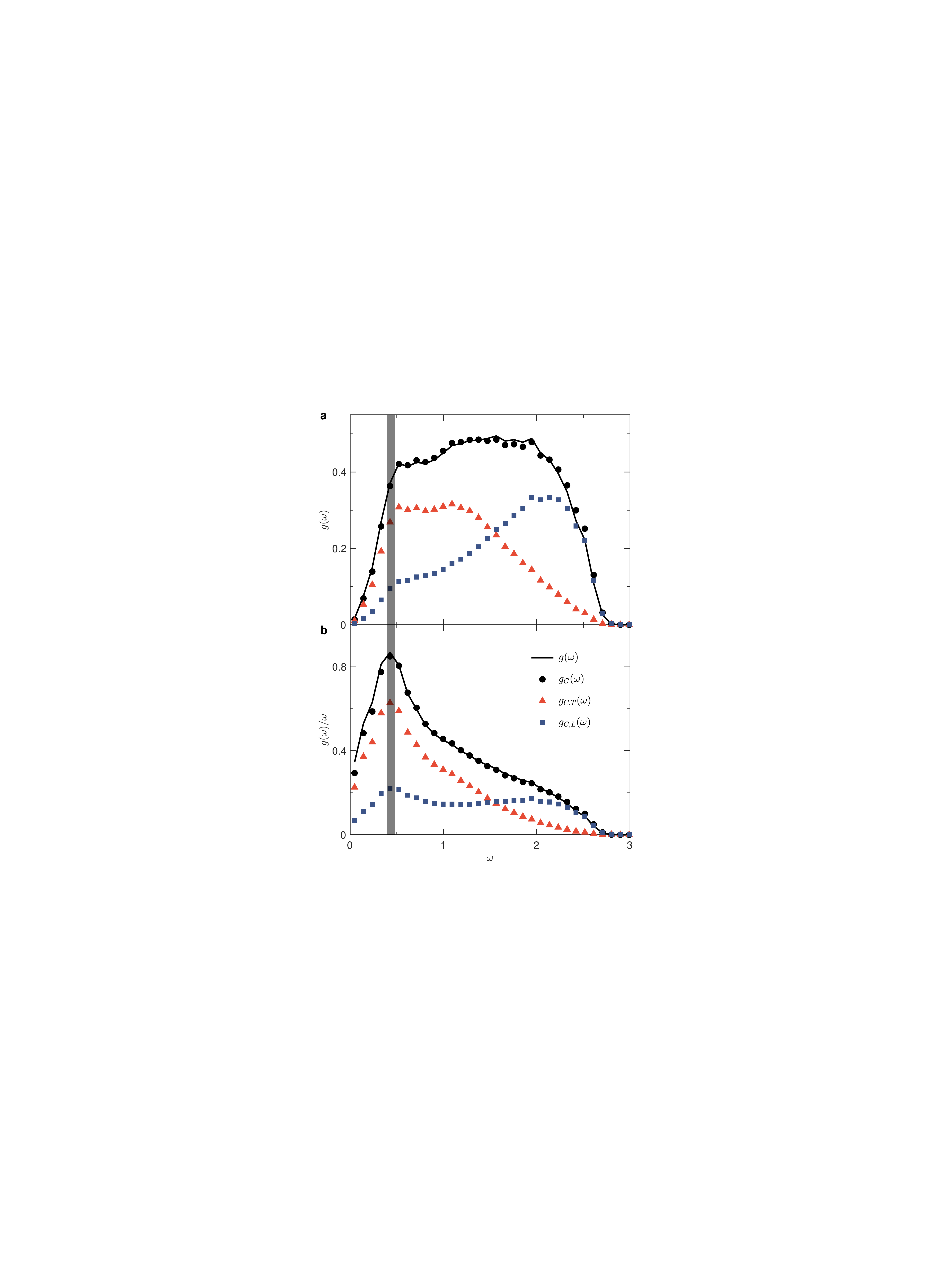}}
	\caption{a, The DOS $g(\omega)$ (the distribution of eigen-frequencies) is plotted using a black solid curve. The black solid circles represent $g_C (\omega)$, which is an integration of the current-correlation functions from 0 up to 1.1$\times$(Debye wavenumber) and is decomposed into the transverse (orange triangles) and longitudinal (blue squares) components. b, The reduced DOS $g(\omega)/\omega$. The gray thick line marks the boson peak position.}
	\label{Fig2}
\end{figure}

In Fig. \ref{Fig2}, we show the obtained DOS $g(\omega)$
and the reduced
DOS $g(\omega)/\omega$, with a BP at $\omega_{BP}\approx 0.43 \approx 0.18\omega_D$ \cite{debye}, i.e. an enhancement over the Debye DOS, for which the reduced DOS would be constant.
Here, the frequency $\omega$ is in units of $\sqrt{\langle k_n \rangle /m}$, where $\langle k_n \rangle$ is the average normal spring constant.

From the eigenvectors we calculate the transverse and longitudinal current-correlation functions (CCFs)
\begin{equation}\label{ccf1}
C_T(k,\omega)=\sum_{\lambda}\delta(\omega-\omega_{\lambda})
\left|\sum_{\lambda}[\hat{k}\times\mathbf{e}_\lambda(j)]\exp (i \vec{k}\cdot \vec{r}_j)\right |^2
\end{equation}
\begin{equation}\label{ccf2}
C_L(k,\omega)=\sum_{\lambda}\delta(\omega-\omega_{\lambda})
\left|\sum_{\lambda}[\hat{k}\cdot\mathbf{e}_\lambda(j)]\exp (i \vec{k}\cdot \vec{r}_j)\right |^2
\end{equation}
Here $\hat{k}=\vec{k}/|\vec{k}|$,
and $\vec{r}_j$ is the position of disk $j$.

The DOS can be calculated alternatively with the help of the CCFs of Eqs. (\ref{ccf1}) and (\ref{ccf2}) \cite{Schirmacher06,MonacoGiordano09,Schirmacher15},
\begin{equation}\label{dosc}
\begin{split}
g_C(\omega)&=\frac{1}{Nk_{\rm max}^2}\int_0^{k_{\rm max}}k(C_L(k,\omega)+C_T(k,\omega))dk \\
&=g_{C,L}(\omega)+g_{C,T}(\omega).
\end{split}
\end{equation}
where $k_{\rm max}$ should be near the Debye wave vector $k_D$ \cite{debye}.
By this we are able to trace the origin of the eigenfunctions corresponding to the eigenvalues sampled in the DOS. Comparing with Eq. (\ref{dos}) we find agreement between $g(\omega)$ and $g_C(\omega)$ for $k_{\rm max}=1.1k_D$.

From the contributions to $g_C(\omega)$ displayed also in Fig. \ref{Fig2},
the BP is dominated by the transverse modes,
in agreement with theoretical \cite{Schirmacher06,Schirmacher07,Schirmacher15} 
and numerical results \cite{Shintani08, Marruzzo13}.

\begin{figure}
	\centerline{\includegraphics[trim=0cm 0cm 0cm 0cm, width=1.0\linewidth]{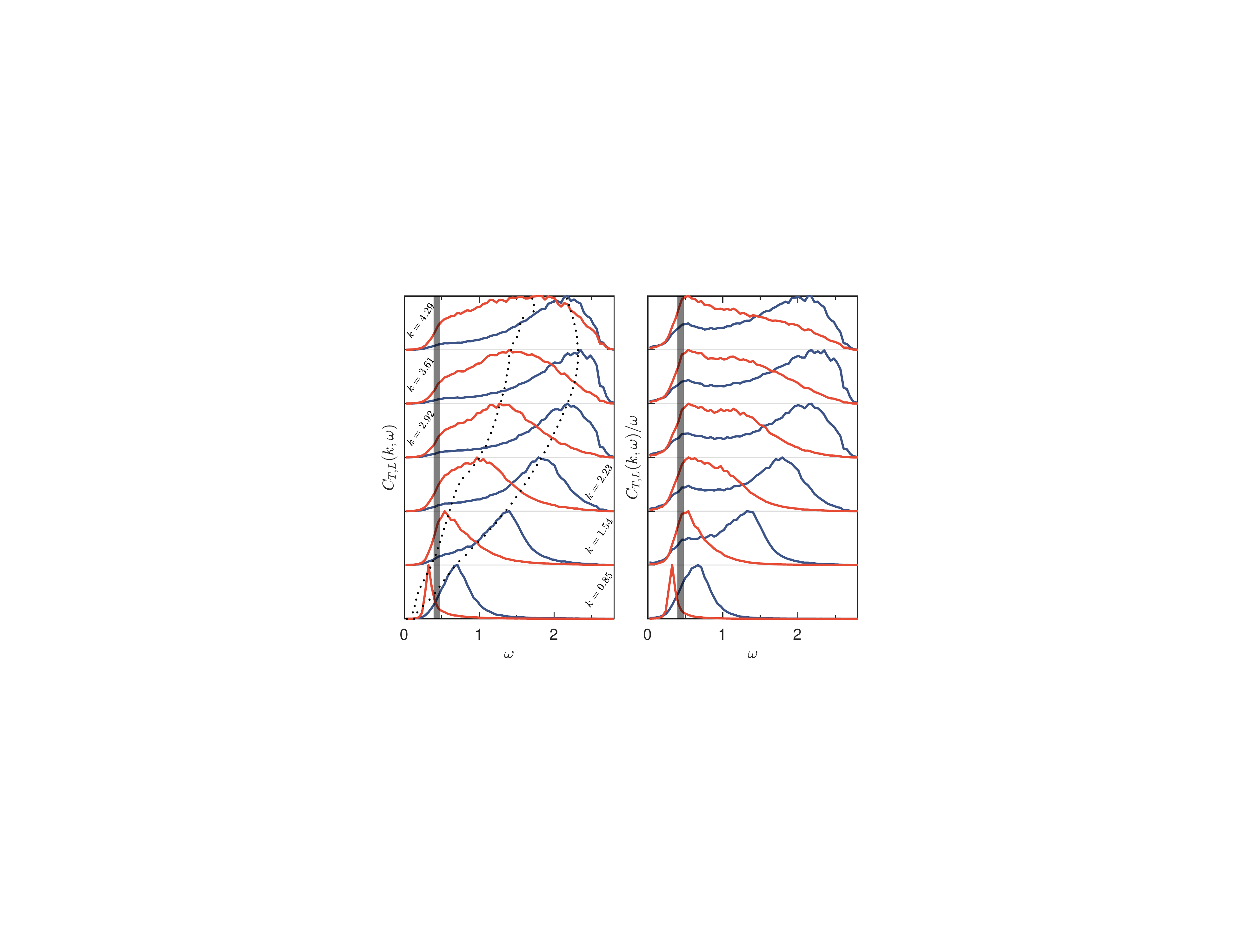}}
	\caption{
		Left: Transverse (orange, left) and longitudinal (blue, right) current-correlation functions (CCFs) $C_{T,L}(k,\omega)$ for $k = 0.85, 1.54, 2.23, 2.92, 3.61, 4.29$ (in units of $\langle D \rangle ^{-1}$) from bottom to top, which are rescaled by their maximum. Dotted lines connect the maximum of $C_{T,L} (k,\omega)$, denoted using $\Omega_{T,L}^{\rm max}(k)$. \newline
		Right: Reduced CCFs $C_{T,L} (k,\omega)/\omega$. }
	\label{Fig3}
\end{figure}

\begin{figure}
	\centerline{\includegraphics[trim=0cm 0cm 0cm 0cm, width=1.0\linewidth]{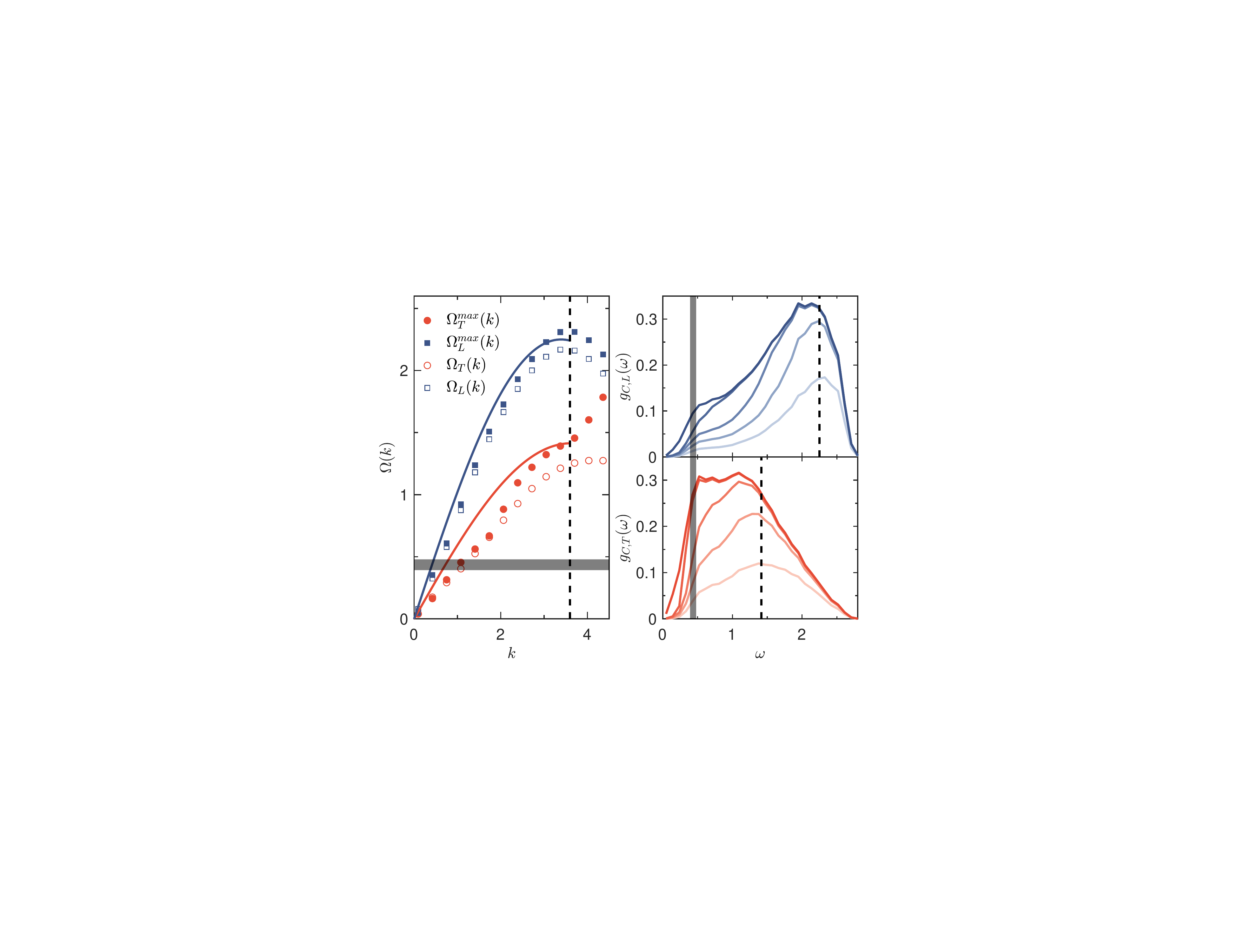}}
	\caption{
		Left: Pseudo-dispersion relations of $\Omega_{T,L}^{\rm max} (k)$ (solid symbols) and $\Omega_{T,L}(k)$ (open symbols), which are obtained from the resonance frequencies in the DHO fitting of the CCFs. The solid curves represent the dispersion relations of a triangular lattice. The vertical dashed line indicates the Debye wavenumber $k_D\approx 3.6$. \newline
		Right: The longitudinal (top) and transverse (bottom) DOS derived from the integration of CCFs from $k_{\rm min}$ to $k_{\rm max}=1.1\times k_D$, with the lower integration limit $k_{\rm min}=[0.8,\ 0.6,\ 0.4,\ 0.2,\ 0]\times k_{\rm max}$, from bottom to top. The vertical dashed lines indicate the first (transverse) and  second (longitudinal) Van Hove singularities of the triangular lattice. The gray thick lines mark the boson peak position.}
	\label{Fig4}
\end{figure}

In Fig. \ref{Fig3}, we plot the $C_{T,L}(k,\omega)$ in the relevant wavenumber $k$ range, as well as the $C_{T,L}(k,\omega)/\omega$, corresponding to the reduced DOS.
In the left panel of Fig. \ref{Fig4}, we plot the maximum, $\Omega^{\rm max}_{T,L}(k)$, of the $C_{T,L}(k,\omega)$
against $k$ versus the transverse and longitudinal dispersions of a regular triangular lattice
$\omega_T=\sqrt{\frac{2K}{m}}\sin(\frac{\sqrt3}{4}k)$ and
$\omega_L=\sqrt{\frac{K}{m}}\sqrt{1-\cos \frac{k}{2}+2(1-\cos k)}$,
where $K$ is the spring constant. 
Clearly, $\Omega^{\rm max}_{T,L}(k)$ follow the crystalline dispersions and level off near the pseudo BZ boundary at $k_D$ \cite{debye}.

In the right panel of Fig. \ref{Fig4}, we show the DOS as evaluated with Eq. (\ref{dosc}), but instead of the lower integral boundary $k_{\rm min}=0$ we used several finite values for $k_{\rm min}$ (see caption).
We see that, when $k_{\rm min}$ approaches the $k$ value where the leveling-off of the longitudinal and transverse dispersions
happens, the peaks in the DOS align with the VHSs of the triangular lattice at $\omega_{VH1}=\sqrt{2}$ and $\omega_{VH2}=2.25$, very near $\Omega^{\rm max}_{T,L}(k)$, whose $k$ is at the pseudo BZ boundary as indicated by the dashed line.

By fitting 
\begin{equation}\label{dho1}
C_{T,L}(k,\omega)\propto \frac{\omega^2}{(\omega^2-\Omega_{T,L}(k)^2)^2+\omega^2\Gamma_{T,L}(k)^2}
\end{equation}
using damped-harmonic-oscillator functions (DHO) (see Refs. \cite{Marruzzo13,Shintani08} and the Appendix), one can
identify intrinsic dispersion functions $\Omega_{T,L}(k)$ and attenuation functions (Brillouin line widths) $\Gamma_{T,L}(k)$. From the left panel of Fig. \ref{Fig4} we can see, that the curves $\Omega_{T,L}(k)$ agree nicely to $\Omega^{\rm max}_{T,L}(k)$, and those of the triangular (crystalline) lattice.

We emphasize, that in our sample there is by no means a triangular long-range order.
However, we verified that on average particles have sixfold coordination as in the triangular lattice by calculating the
integral over the first coordination shell of the radial distribution function $g(r)$, $\bar Z=2\pi\int_0^{R_C}  dr r g(r)\approx$ 5.5 (Here $R_C$ is the first minimum of $g(r)$, see Appendix Fig. 8).
The fact that the ``glassy dispersions'' $\Omega_{L,T}(k)$ agree to the dispersions of the triangular lattice (including the VHS)
is obviously due to the almost six-fold short-range order.

Therefore, we clearly observe what was described by the authors of Refs.
\cite{Chumakov11,%
Chumakov14,%
Chumakov15,%
Chumakov16}
as a would-be scenario for the origin of the BP. However, the VHS, namely
the leveling off of the transverse dispersion occurs at a much
higher frequency, completely separated from $\omega_{\rm BP}$.

\subsection{Anomalies associated with the boson peak}
\begin{figure}
	\centerline{\includegraphics[trim=0cm 0cm 0cm 0cm, width=1\linewidth]{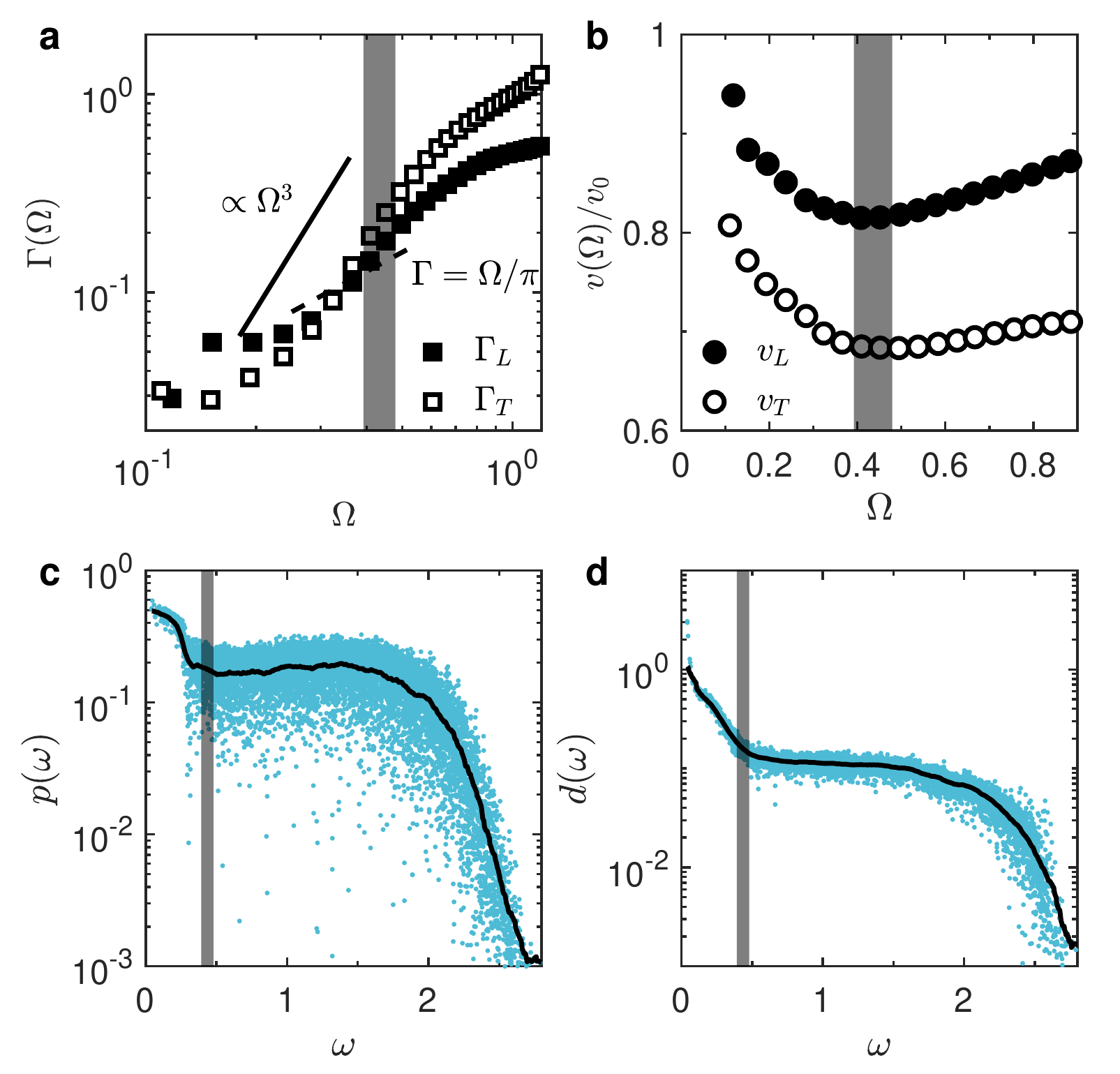}}
	\caption{Frequency dependence of sound attenuations $\Gamma_{T,L}(\Omega)$ (a) and sound velocities $v_{T,L}(\Omega)=\Omega_{T,L}(k)/k$, normalized by macroscopic velocities $v^{T,L}_0$ (b) for longitudinal and transverse modes. Frequency dependence of the participation ratio (c) and the diffusivity (d). The gray thick lines in all panels mark the boson peak position. In panel (a), the black solid line $\Omega^3$ is a guide to the eye. The Ioffe-Regel limits of the transverse and longitudinal modes are given by the crossing points between the dashed line $\Omega/\pi$ and the curves of $\Gamma_{T,L}(\Omega)$, which occur slightly before the boson peak position. In panels (c,d), the dots represent individual modes and the solid curves are the average.}
	\label{Fig5}
\end{figure}

As we now see that the BP is {\it not} identical to the VHS, in contrast to 
Refs.
\cite{Chumakov11,%
Chumakov14,%
Chumakov15,%
Chumakov16},
we now analyze in detail the vibrational states giving rise to the BP in terms of the structural disorder. 
One prominent feature is the existence of a disorder-induced sound attenuation $\Gamma_{T,L}$,
corresponding to the Brillouin line width of inelastic neutron and x-ray scattering (Brillouin scattering) experiments \cite{Baldi10,MonacoGiordano09},
as evaluated in the DHO fits and plotted in Fig. \ref{Fig5}(a). In panel (b), we plot
sound velocities $v_{T,L}(\Omega)=\Omega_{T,L}(k)/k$,
rescaled by the macroscopic velocities
$v_0^T=\sqrt{G/\rho}$,
$v_0^L=\sqrt{(G+B)/\rho}$.
We see a characteristic dip in $v_{T,L}(\Omega)$ just near $\omega_{BP}$, where $\Gamma(\Omega)$ is steepest.

Figures \ref{Fig5}(a,b) agree with the heterogeneous elasticity theory \cite{Schirmacher07,Schirmacher15},
where the elastic-constant disorder produces {\it frequency-dependent complex elastic moduli}.
For the transverse elastic modulus, we have
\begin{equation}\label{vt1}
\hat G(\omega)=
G'(\omega)-iG''(\omega)
=G'(\omega)[1-i\Gamma_T(\omega)/\omega].
\end{equation}
Near the Brillouin resonance, we may write
\begin{equation}\label{vt2}
\hat G(\omega)\approx\hat G\big(\Omega\big)
=G'(\Omega)[1-i\Gamma_T\big(\Omega\big)/\Omega].
\end{equation}
So 
$v_T(\Omega)^2\propto G'\big(\Omega\big)$ and
$\Gamma_T(\Omega)\propto \Omega G''\big(\Omega)$ are related to each other
by the Kramers-Kronig transformation, as dictated by
causality \cite{Schirmacher15}, meaning where $G''(\Omega)$ has its
strongest increase $G'(\Omega)$ must have a dip.
As shown by Ref. \cite{Schirmacher07}, the BP is produced
by the disorder-induced strong increase of $\Gamma_{T}(\Omega)$.
So the BP, the strong increase of $\Gamma_{T}(\Omega)$, and
the dip in $v_T(\Omega)$ are just the same phenomenon.
These three anomalies come about, because on the length scale of $L_{BP}=2\pi/k_{BP}$, comparable to the spatial extent of the
elastic-constant fluctuations \cite{Schirmacher15,Wyart14}, the system is no more effectively homogeneous and isotropic
(as it is for large scales). At this length scale, which is about 6 `atomic' (disk) diameters, the vibrational excitations are
no more Debye-type plane waves but random-matrix-type modes,
which cannot be degenerate because of the absence of symmetries.
The strong increase of $\Gamma(\Omega)$ near $\omega_{BP}$ in many cases
follows a $\Gamma(\Omega)\propto\Omega^{d+1}$ law (Rayleigh scattering) \cite{Baldi10,MonacoGiordano09,MonacoMossa09,Marruzzo13}.
Indeed, in our case, $\Gamma_T(\Omega)$  is compatible to $\Omega^3$ just below the BP,
as depicted in Fig. \ref{Fig5}(a); in addition, the Ioffe-Regel limit of both the transverse and longitudinal modes occurs slightly below, $\omega_{BP}$, consistent with Refs. \cite{Shintani08, Baldi10, Marruzzo13}, suggesting that acoustic modes stop propagation and become diffusive near the BP.

Further evidence for the disorder-induced nature of the BP comes from considering the participation ratio $p(\omega)$  and the frequency-dependent diffusivity $d(\omega)$ \cite{Xu09}.
The participation ratio
$
p(\omega_{\lambda})=
\left.
\left(\sum_i \left| \mathbf{e}_{\lambda}(i)\right |^2\right)^2
\middle /
\left(N\sum_i \left| \mathbf{e}_{\lambda}(i)\right|^4\right)
\right.
$
is expected to be comparable to 1 for de-localized plane-wave-like states and of the order of $1/N$ for Anderson-localized states, i.e. states which are only finite in a certain region.
In Fig. \ref{Fig5}(c), $\omega_{BP}$ marks a crossover frequency, around which $p(\omega)$ stops decreasing and reaches a plateau; eventually at sufficiently high frequency $\omega_{\rm loc}\approx2.16$, $p(\omega)$ drops sharply to a value comparable to $1/N$ indicating Anderson-localization near $\omega_D$, in agreement with the literature (e.g. \cite{Schirmacher98}).

We confirm this scenario by considering the frequency dependent diffusivity $d(\omega)$ \cite{Xu09}.
Below $\omega_{BP}$, $d(\omega)$ decreases as is typical for the Debye wave regime.
Between $\omega_{BP}$ and $\omega_{\rm loc}$ the diffusivity is finite and constant, so $\omega_{BP}$ marks the crossover between the nearly-free wave and diffusive regimes, in agreement with the crossing of the Ioffe-Regel limit for the transverse excitations near $\omega_{BP}$ as seen in Fig. \ref{Fig5}(a).

\subsection{Structural signatures associated with the boson peak}
\begin{figure}
	\centerline{\includegraphics[trim=0cm 0cm 0cm 0cm, width=1\linewidth]{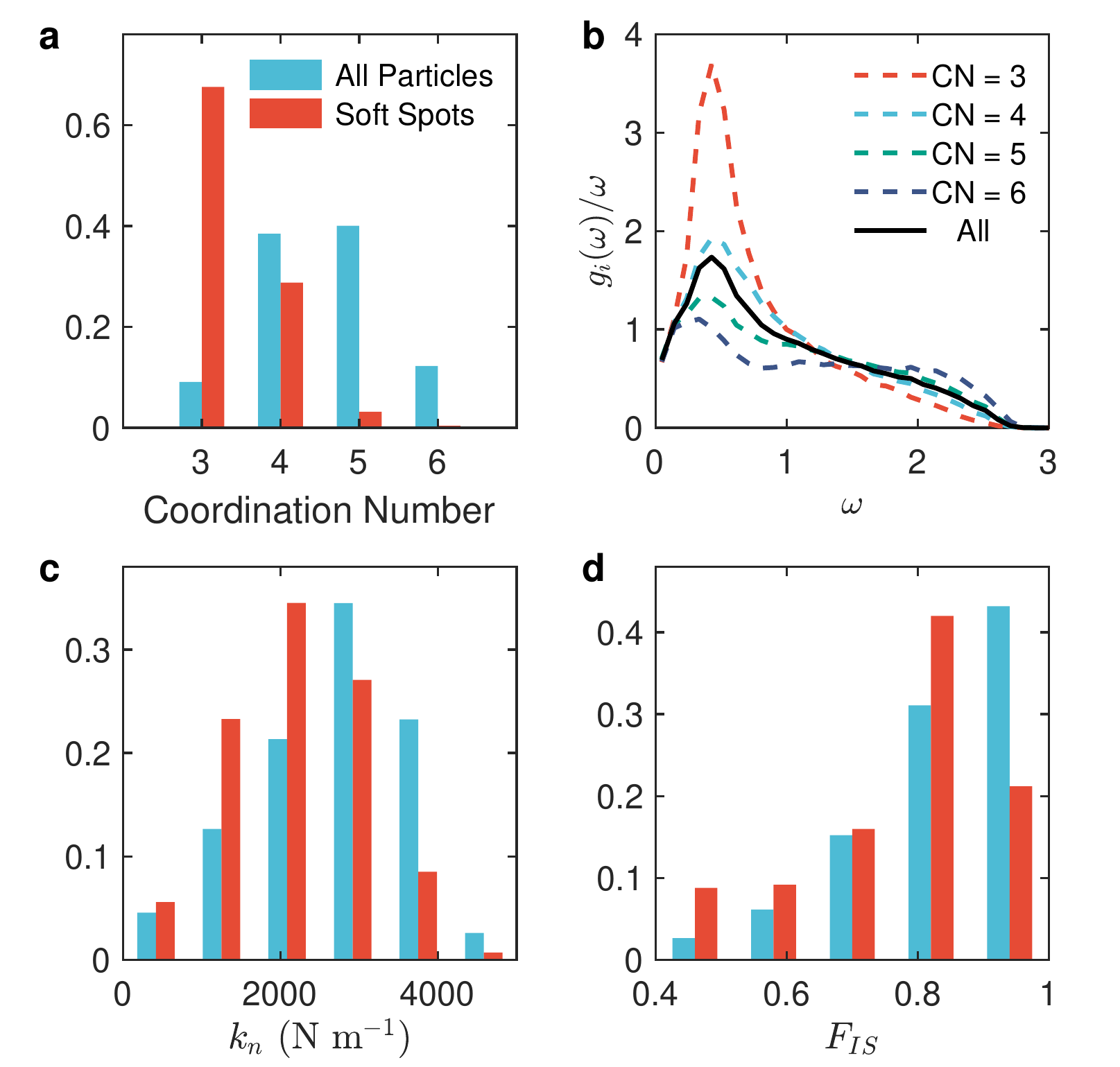}}
	\caption{Probability distributions of contact coordination number (a), spring constant (c) and local inversion-symmetry parameter (d) for all particles (left blue columns) and soft spots (right red columns). b, Reduced density of states per particles $g_i (\omega)/\omega$ averaged over particles of different coordination numbers, and the black solid curve represents $g_i (\omega)/\omega$ averaged for all particles.}
	\label{Fig6}
\end{figure}

We further substantiate that the BP is related to so-called ``soft spots" in our sample,
which have been investigated recently in connection with the plastic movement of glasses under shear \cite{Tanguy10,Manning11}.
We define soft spots in the following way: We consider the statistics of local vibrational intensities $|e_\lambda(i)|^2$ near $\omega_{BP}$. Sites $(i)$, which belong to the top 5 \% of the statistics, are called ``soft spots''.

In Fig. \ref{Fig6}, we compare the spectral statistics of all sites with those of soft spots for (a) the contact coordination number (CCN), (b) the reduced single-site DOS $g_i(\omega)/\omega$, (c) the average magnitude of spring constants,
and (d) the local inversion-symmetry parameter $F_{IS}$, as introduced by Zaccone \cite{Milkus16}.
The CCN of soft spots have a significant contribution from $\rm CCN=3$, whereas they distribute symmetrically around $\rm CCN = 4.5$ for all sites.
For $g_i (\omega)/\omega$, particles of $\rm CCN=3$ make significant contributions to the BP. Moreover, the distribution of spring constants of soft spots shifts down compared to that of all particles, as shown in Fig. \ref{Fig6}c. The $F_{IS}$, which is unity for crystals and decreases as the central symmetry breaks down, appears lower for soft spots than for all particles, consistent with some recent ideas \cite{Milkus16}.

\section{Discussion}
Let us now discuss the relevance of our findings with the boson-peak related vibrational anomalies in three-dimensional
real glassy materials.

These anomalies have been identified by spectroscopic methods namely Raman scattering \cite{raman}, as well as
inelastic neutron, x-ray \cite{MonacoGiordano09,Baldi10} and nuclear \cite{chumakov04} scattering.
As the scattered intensity followed the temperature dependence of the boson occupation function, (from which the name ``boson peak'' was coined) the conclusion was that the fluctuation spectrum of the excitation was temperature independent, pointing to harmonic degrees of freedom.
So the discussion concentrated on characterizing the dynamical matrix of glasses, in order to relate the glass structure to the observed anomalies. 

As mentioned above, these efforts led to conflicting characterizations of the boson peak-related anomalies
in terms of elastic disorder (heterogeneous elasticity) \cite{Karpov83,Elliott92,Gurevich93,Schirmacher98,Schirmacher06,Schirmacher07,Marruzzo13}, as well as in terms of washed-out Van Hove singularities, created by short-range order \cite{Chumakov11,Chumakov14,Chumakov15,Chumakov16}.

Model systems with repulsive soft-sphere interactions proved in the past to be very useful for characterizing and understanding
the vibrational features of glasses \cite{schober93,schober04,Marruzzo13}.

Our soft-disc model system (which is not a virtual but a real one) serves as an analog simulation of a disordered dynamical matrix
of a glass. 
We find both evidence for a crystal-like dispersion leading to Van Hove-singularity-like features in the density of states, as well as evidence for vibrational anomalies as characterized by heterogeneous-elasticity theory.

What makes our model system different from glassy materials,
namely that it is macroscopic and two-dimensional,
is, in fact, not a disadvantage, but, on the contrary,
serves to disentangle the disorder-related and the short-range-order related features of glasses.

Obviously in many glasses, especially those investigated by the advocates of the Van Hove-singularity model, the scale of the molecular units and the range of the disorder correlations are approximately the same, so that it is difficult to separate the spectroscopic
consequence of structural disorder and short-range order.

In our system, these scales are almost one order of magnitude different, leading to a clear separation of the Van Hove singularities and the disorder-induced boson peak.

\section{Conclusions}
Our findings can be summarized as follows:

We have carefully prepared a 2d model glass, which has a completely amorphous structure but still predominantly sixfold nearest-neighbor coordination. 
By evaluating the current-correlation functions, we observe a bending down of the transverse and longitudinal dispersions
of the vibrational excitations near the pseudo-Brillouin-zone radius $k_p\approx k_D$. This bending down leads to a piling up
of vibrational states as in the Van Hove singularities of crystals and leads to corresponding maxima in the density of states near
the Van Hove singularities of the triangular lattice. Such a scenario was made responsible for the appearance of the boson peak in glasses by Chumakov et al.
\cite{Chumakov11,%
Chumakov14,%
Chumakov15,%
Chumakov16}

However, we observe a boson peak, i.e. a peak in the reduced
density of states at a much lower frequency as that of the transverse Van Hove
singularity.

The boson peak shows all the features of a disorder-induced enhancement of the density of states as described by heterogeneous elasticity theory \cite{Schirmacher06,Schirmacher07,Schirmacher15}: The boson-peak frequency coincides with the Ioffe-Regel frequency and marks the transition from a weakly-damped wave regime to the regime of diffusive wave transport.
The boson-peak wavenumber $k_{\rm BP}=\omega_{\rm BP}/v_T=2\pi/L_{\rm BP}$ denotes the length scale at which the waves start to feel the breakdown of the continuum symmetry.

With the help of our model system we hope to have made clear that the washed-out Van Hove singularities and the boson peak in glasses are two separate physical phenomena. The former are a result of the short-range order, reminiscent of the crystalline state. The latter is a result of the structural disorder, produced by the breakdown of the continuum symmetry near the boson-peak length scale.
While both features are likely to occur in glasses, the coincidence of the Van Hove and boson-peak length scales, observed
in some materials, does not mean that the two phenomena are the same.

The features accompanying the disorder-related boson peak, namely the rapid increase of the attenuation and the characteristic
dip in the group velocity are a way to disentangle the boson peak and the Van Hove features in real glass. In our model glass
the two features are separated due to our special geometry.

\section*{Acknowledgments}

J. Z. acknowledges support from the NSFC under Awards No. 11474196 and No. 11774221.\\

\section*{Appendix: details of the structural and spectral analysis}
\begin{figure}
	\centerline{\includegraphics[trim=0cm 0cm 0cm 0cm, width=1\linewidth]{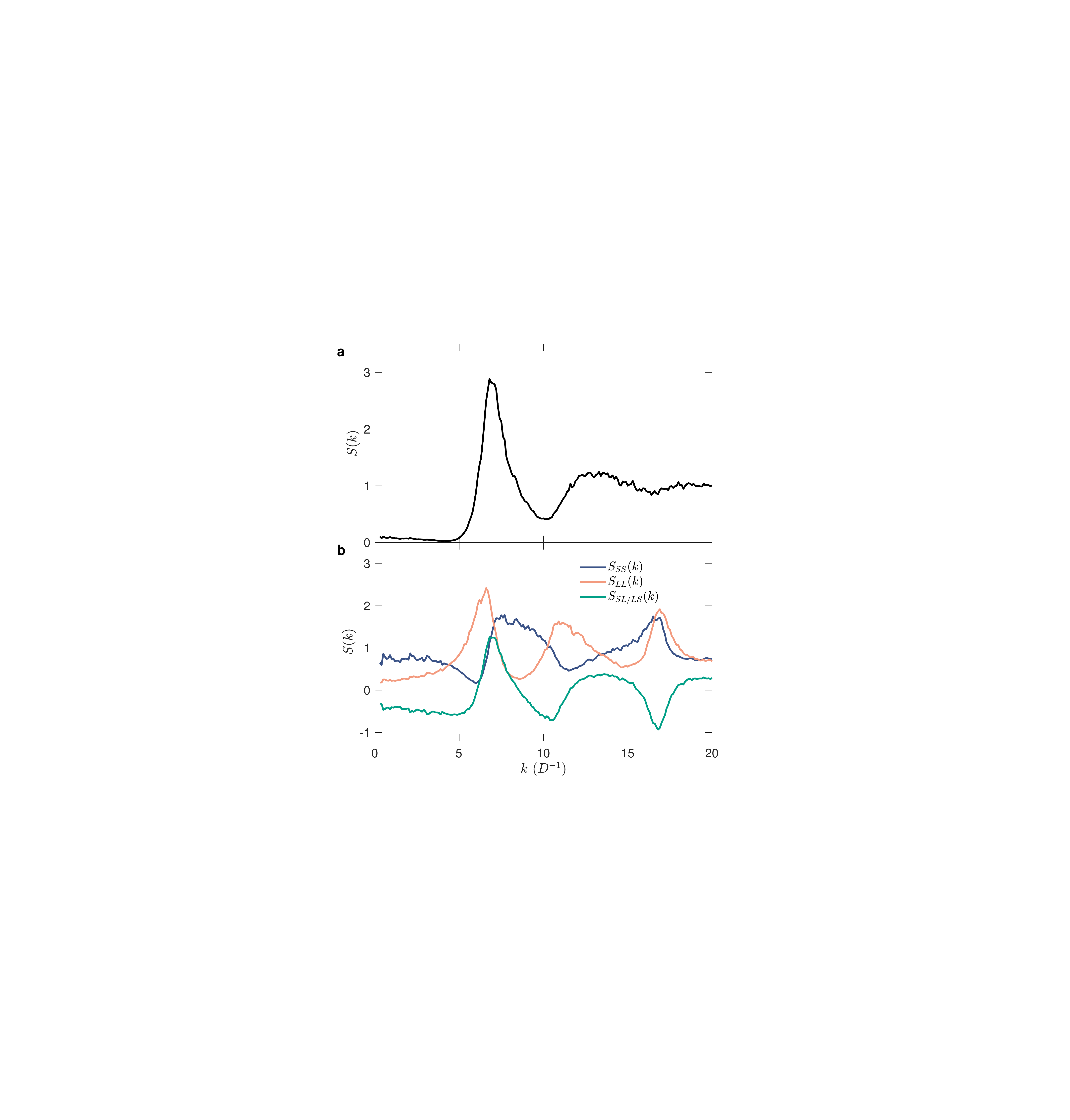}}
	\caption{(a) Static structure factors $S(k)$ for all particles. (b) Partial static structure factors (S: small disks, L: large disks).}
	\label{Fig7}
\end{figure}
  
In Fig. \ref{Fig7}, we show the static structure factor
\begin{equation*}
S(k)=\frac{1}{N}\sum_{i=1}^N\sum_{j\neq i}\left\langle
e^{-i\bm{k}(\bm{r}_i-\bm{r_j})}\right\rangle
\end{equation*}
where $\bm{r}_i$ are the positions of the centers of the disks
and $N$ the total number ($N$ = 5440) of the disks.

\begin{figure}
	\centerline{\includegraphics[trim=0cm 0cm 0cm 0cm, width=1\linewidth]{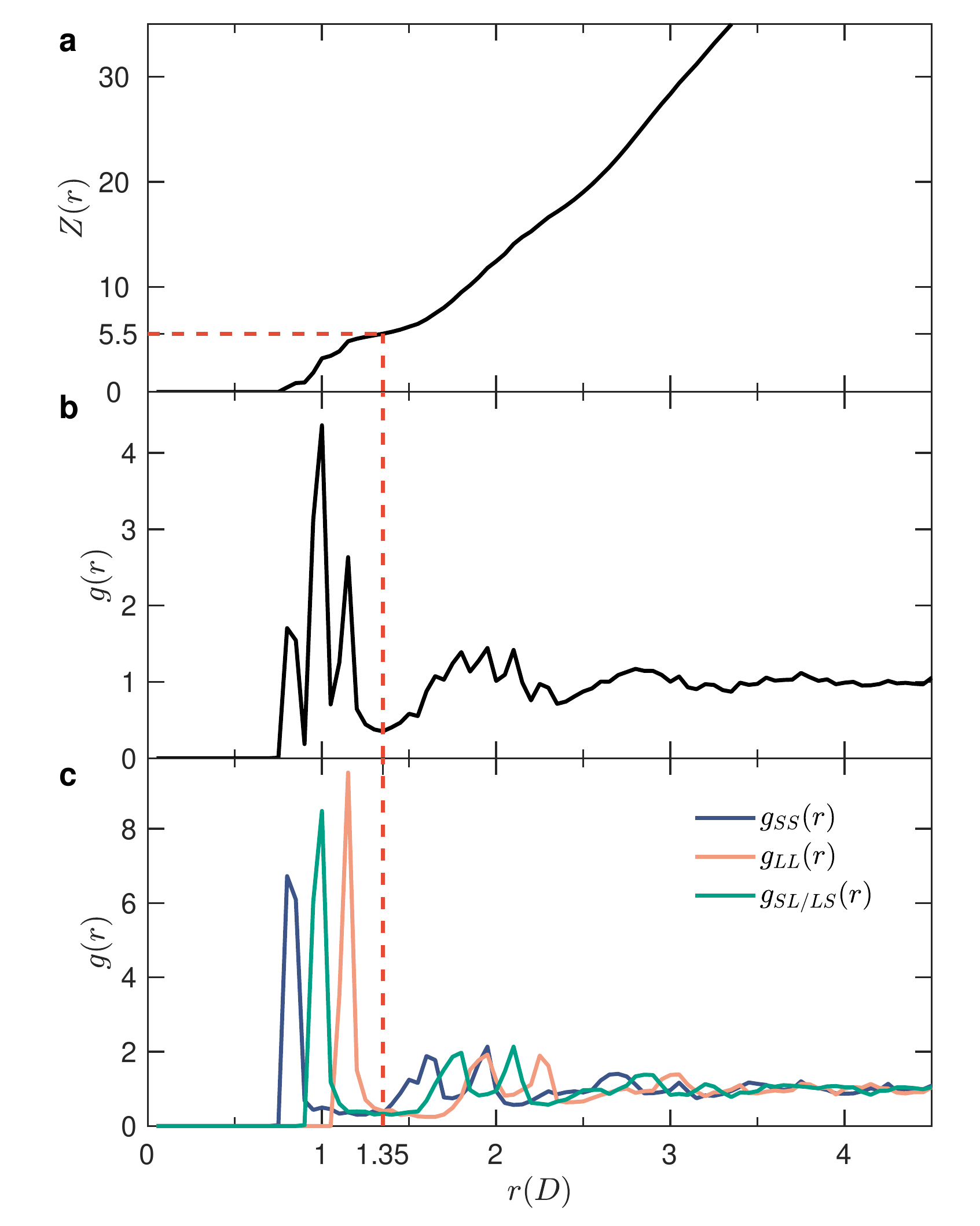}}
	\caption{Integrated radial distribution function $Z(r)$, radial distribution function $g(r)$ and partial radial distribution function (S: small disks, L: large disks).}
	\label{Fig8}
\end{figure}

In Fig. \ref{Fig8}b, we show the radial distribution function
\begin{equation*}
g(r) = \frac{1}{N\rho}\sum_{i=1}^{N}\sum_{j\neq i}\langle \delta (\bm{r}+\bm{r}_i-\bm{r}_j)\rangle
\end{equation*}
$\rho =N/A$ is the density, $A$ is the area of the biax.
In the upper panel we show the integrated radial distribution function
\begin{equation*}
Z(r)=2\pi\int_0^rd\tilde r\tilde rg(\tilde r)
\end{equation*}
which can be interpreted as an $r$ dependent coordination number.
$Z(r)$ gives the number of disks around a given disks the center of
which has a distance from the given one smaller or equal to $r$.
We see that this function has a plateau where $g(r)$ has a broad minimum
at $r_{\rm min}=1.35$.
This minimum defines the first coordination shell. This leads to a
value of $Z(r_{\rm min})=$ 5.5.

\begin{figure}
	\centerline{\includegraphics[trim=0cm 0cm 0cm 0cm, width=1\linewidth]{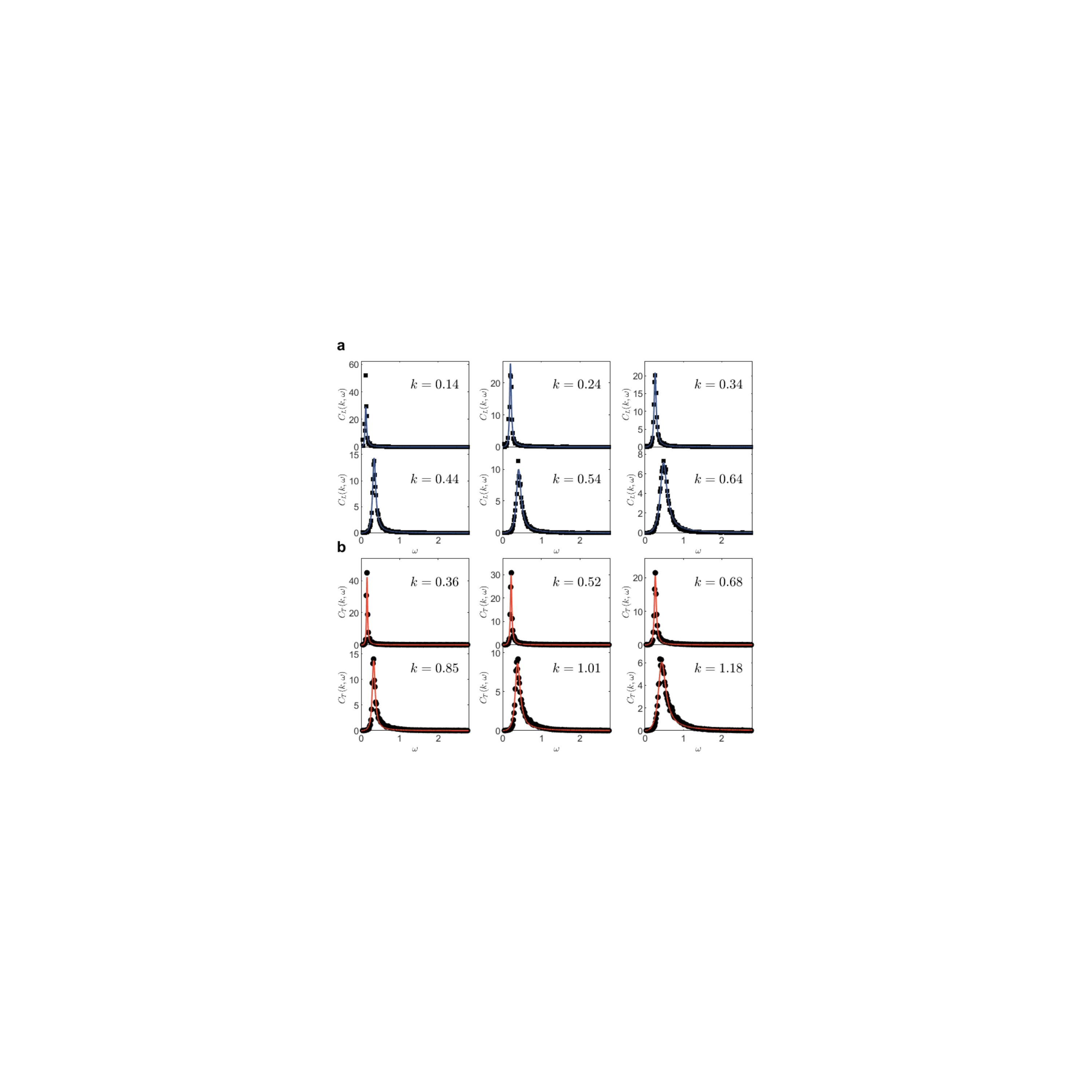}}
	\caption{Longitudinal (a) and transverse (b) current-correlation functions and DHO fitting (solid lines).}
	\label{Fig9}
\end{figure}

In Fig. \ref{Fig9}, we show the current-current correlation functions
fitted with the damped-harmonic oscillator (DHO) function
\begin{eqnarray}\label{dho2}
C_{L,T}(k,\omega)&\propto&
\omega
\mbox{Im}\{
\frac{1}{
	\Omega_{L,T}^2(k)-\omega^2-i\omega\Gamma_{L,T}(k)}
\}\qquad\nonumber\\
&=&
\omega^2
\frac{\scriptstyle
	\Gamma_{L,T}(k)
}{\scriptstyle
	(\Omega_{L,T}(k)^2-\omega^2)^2+
	\omega^2\Gamma_{L,T}(k)^2
}\nonumber
\end{eqnarray}


\begin{thebibliography}{36}
\bibitem[{deb()}]{debye}
\bibinfo{note}{$\omega_D=v_Dk_D$ is the Debye frequency, with the Debye
  velocity $v_D^{-1}=\frac{1}{2}[v_L^{-1}+v_T^{-1}]$ and the Debye wavenumber
  $k_D=\sqrt{4\pi N/A}$. $v_{L,T}$ are the longitudinal and transverse sound
  velocities, $N$ is the number of disks, and $A$ is the sample area. $k_D$ can
  also be identified with the radius $k_p=k_0/2$ of the ``pseudo Brillouin
  zone'' of the amorphous material, where $k_0$ is the first sharp diffraction
  peak in the structure factor. For our sample $k_D=3.6/\langle D\rangle$ and
  $k_0=7.0/\langle D\rangle$, where $\langle D\rangle$ is the average disk
  diameter.}
\end{thebibliography}

\end{document}